\documentstyle[preprint,tighten,aps,psfig]{revtex}

\begin{document}

\preprint{BNL-HET-98/11, hep-ph/9804252}

\title{\boldmath Electroweak radiative corrections to $b\to s\gamma$}

\author{Andrzej Czarnecki and William J. Marciano}
\address{Physics Department, Brookhaven National Laboratory\\
Upton, NY 11973}

\maketitle

\begin{abstract}
Two loop electroweak corrections to $b\to s\gamma$ decays are
computed. Fermion and photonic loop effects are found to reduce
$R\equiv {\cal B}(b\to s\gamma)/{\cal B}(b\to ce\nu)$ by $\sim 8\pm
2\%$ and lead to the standard model prediction ${\cal B} (B\to
X_s\gamma)=3.28 \pm 0.30\times 10^{-4}$ for inclusive $B$ meson
decays.  Comparison of $R^{\rm theory} =3.04\pm 0.25\times
10^{-3}(1+0.10\rho)$, where $\rho$ is a Wolfenstein CKM parameter,
with the current experimental average $R^{\rm exp}=2.52\pm 0.52\times
10^{-3}$ gives $\rho=-1.7\pm 1.9$ which is consistent with $-0.21\le
\rho \le 0.27$ obtained from other $B$ and $K$ physics constraints.
\end{abstract}

\begin{figure} 
\hspace*{-36mm}
\begin{minipage}{16.cm}
\vspace*{3mm}
\[
\hspace*{-10mm}
\mbox{ 
\begin{tabular}{cc}
\psfig{figure=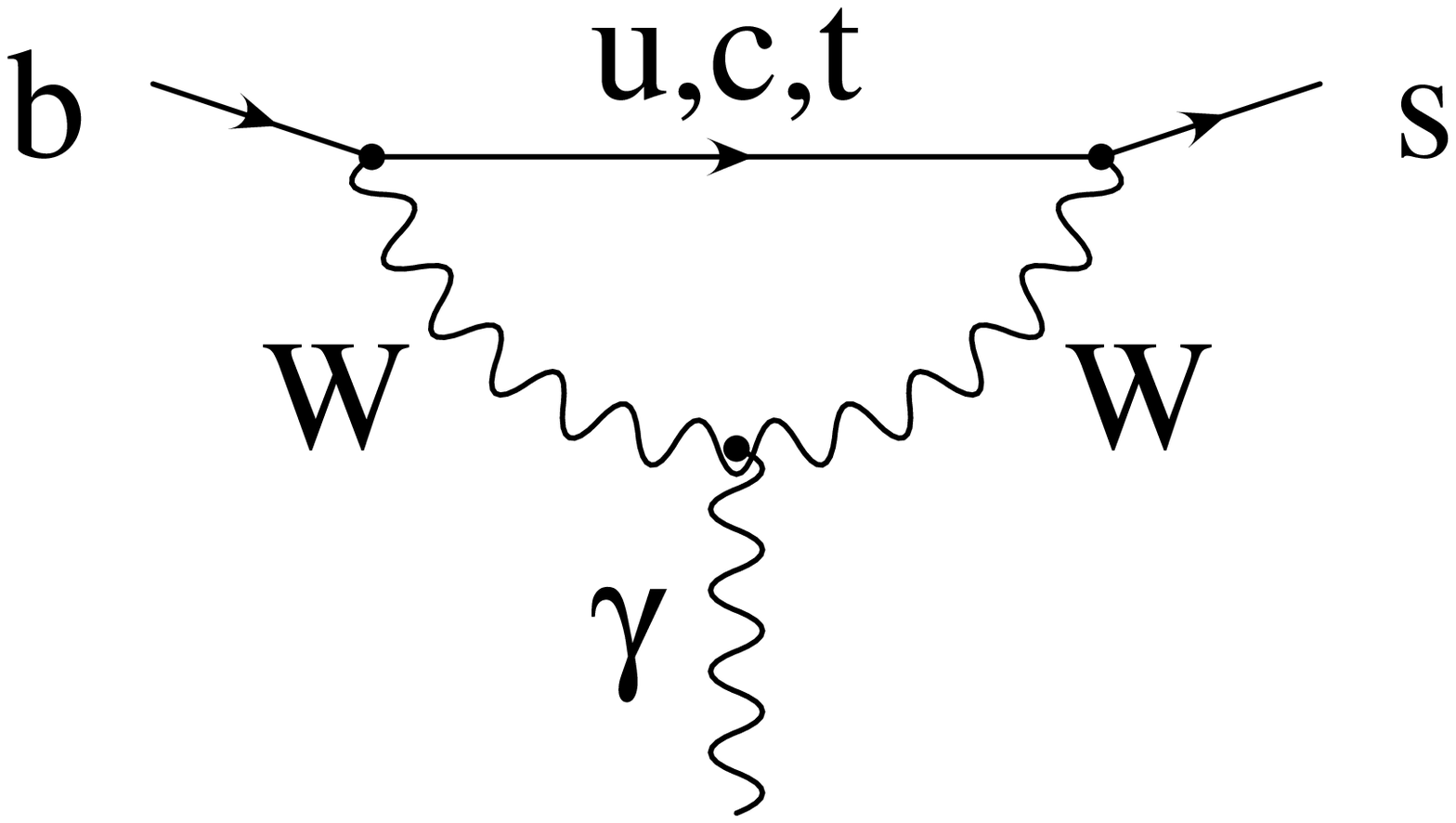,width=35mm,bbllx=72pt,bblly=291pt,%
bburx=544pt,bbury=540pt} 
& \hspace*{8mm}
\psfig{figure=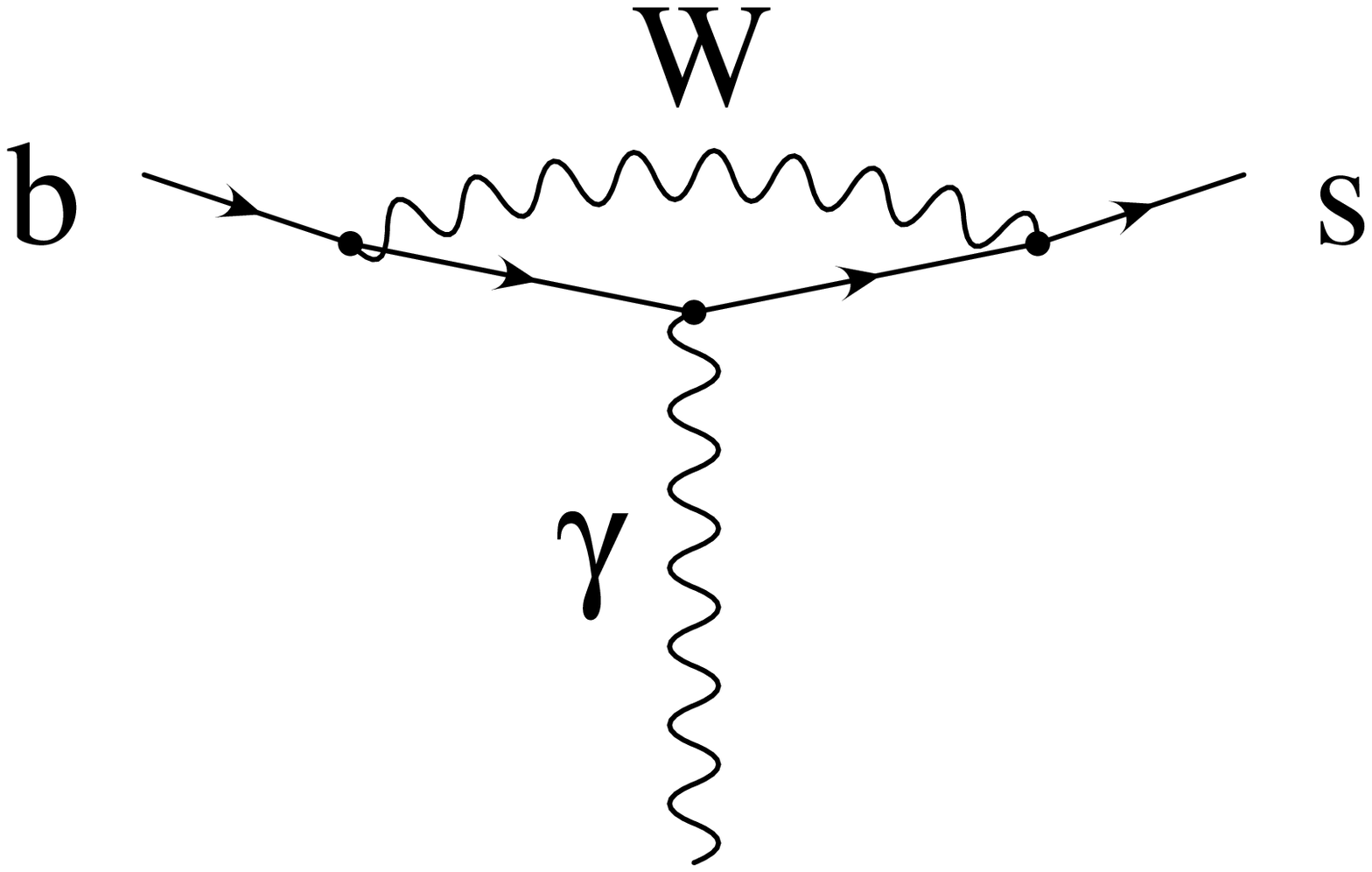,width=35mm,bbllx=72pt,bblly=291pt,%
bburx=544pt,bbury=540pt} 
\end{tabular}
}
\]
\end{minipage}
\caption{One loop diagrams which give rise to the radiative decay
$b\to s\gamma$.} 
\label{fig1}
\end{figure}

\vspace*{0.2cm} 
The inclusive radiative decay $b\to s\gamma$ provides
a quantum loop test of the Standard Model and sensitive probe of new
physics \cite{Campbell:1982rg}. That flavor changing neutral current
reaction proceeds via virtual top, charm, and up quark penguin
diagrams (see fig.~\ref{fig1}).  As such, it depends on and can
provide a determination of the quark mixing parameters
$|V_{ts}^*V_{tb}|$ as well as a test of CKM unitarity. Alternatively,
a clear deviation from expectations could suggest evidence for
additional ``new physics'' loops \cite{Hewett93} from supersymmetry,
charged Higgs scalars, anomalous $W$ couplings, etc.

The CLEO collaboration has published a branching ratio
\begin{eqnarray}
{\cal B} (B\to X_s\gamma) = 2.32 \pm 0.57 \pm 0.35 \times 10^{-4} 
\qquad ({\rm CLEO}),
\label{eq1}
\end{eqnarray}
and reported a preliminary update $2.50\pm 0.47\pm 0.39$ (CLEO II),
for inclusive radiative $B$ meson decays produced at the
$\Upsilon(4S)$ resonance (generally assumed to be a 50-50 mixture of
$B^\pm$ and $B^0$ ($\bar{B}^0$)) \cite{Alam:1995aw}. 
The ALEPH collaboration has
reported \cite{Colrain97}
\begin{eqnarray}
{\cal B} (h_b \to X_s\gamma) = 3.11 \pm 0.80 \pm 0.72 \times 10^{-4} 
\qquad ({\rm ALEPH})
\label{eq2}
\end{eqnarray}
for inclusive $h_b=b$ hadrons (mesons and baryons) originating from
$Z$ decays.  These two measurements are quite different and should not
be expected to give identical results. A better ratio for comparing
distinct experiments and theory is \cite{Chetyrkin:1997vx}
\begin{eqnarray}
R\equiv {\Gamma(b\to s\gamma)\over \Gamma(b\to ce\nu)}
={{\cal B}(b\to s\gamma)\over {\cal B}(b\to ce\nu)}
\simeq 0.96 {{\cal B} (B\to X_s\gamma) \over {\cal B}(B\to X_c e\nu)}
\label{eq3}
\end{eqnarray}
where the 0.96  is a non-perturbative conversion factor specific
to $B$ meson decays which incorporates $1/m_b^2$ and $1/m_c^2$
corrections \cite{Voloshin:1997gw}
We do not know the analogous conversion factor for $h_b$ decays at the
$Z$ pole, but assume it is not very different since $h_b$ consists of
about 75\% $B$ ($B_u$ and $B_d$) mesons.

$R$ should be less sensitive to the $b$ parentage and other systematic
uncertainties, as long as numerator and denominator have a common
experimental origin. Also, from a theoretical perspective, $R$ is less
sensitive to $b$ quark mass uncertainties which largely cancel, modulo
QCD corrections, in the ratio \cite{Chetyrkin:1997vx}.

Employing (after subtracting out a small $b\to u e \nu$ component)
\begin{eqnarray}
{\cal B} (B\to X_c e\nu) &=& 10.34\pm 0.46 \% 
\qquad \mbox{(at $\Upsilon(4S)$) \cite{Barish:1996cx}}
\label{eq4}
\\
{\cal B} (h_b\to X_c e\nu) &=& 10.96\pm 0.20 \% 
\qquad \mbox{(at $Z$ pole)       \cite{Feindt:1998ps}}
\label{eq5}
\end{eqnarray}
leads to
\begin{eqnarray}
R&=& 2.42\pm 0.59 \times 10^{-3} \qquad\mbox{(CLEO II)}
\nonumber \\
R&=& 2.84\pm 1.06 \times 10^{-3} \qquad\mbox{(ALEPH)}
\label{eq6}
\end{eqnarray}
where all errors have been added in quadrature. (We cannot explain why
the leptonic branching ratio in (\ref{eq5}) appears to be somewhat
larger than (\ref{eq4}). It may be indicative of an underlying
systematic normalization uncertainty \cite{Feindt:1998ps}.)

The results in (\ref{eq6}) are consistent and can be combined to give 
\begin{eqnarray}
R^{\rm exp} = 2.52\pm 0.52\times 10^{-3}.
\label{eq7}
\end{eqnarray}
That finding should be compared with the QCD refined Standard Model
prediction
\begin{eqnarray}
R^{\rm QCD} = {|V_{ts}^*V_{tb}|^2\over |V_{cb}|^2} {6\alpha_{em}F\over
\pi g(m_c^2/m_b^2) }
|D^{\rm eff}|^2
\label{eq8}
\end{eqnarray}
where we employ the notation and results of Chetyrkin, Misiak, and
M\"unz \cite{Chetyrkin:1997vx}.  In that expression, $g$ is a
semileptonic decay phase space factor, $F$ is a QCD correction, and
$D^{\rm eff}$ is an effective $b\to s\gamma$ amplitude coupling
(including gluon bremsstrahlung) corrected up to NLO in QCD
\cite{Chetyrkin:1997vx,Ali:1991tj,Adel:1993ah,%
Buras:1994xp,Grinstein:1990tj,Greub:1996tg,Ciuchini:1994fk} (our
$|D^{\rm eff}|^2$ corresponds to $|D|^2+A$ in
ref.~\cite{Chetyrkin:1997vx,Buras:1997bk,Ciuchini:1997xe}).  We employ
an update of $|D^{\rm eff}|^2$ consistent with the results in
ref.~\cite{Buras:1997bk,Ciuchini:1997xe}.

The electromagnetic coupling $\alpha_{em}=e^2(\mu)/4\pi$ in
(\ref{eq8}) was renormalized at a short-distance scale $\mu_W$,
$m_b<\mu_W<m_W$ by the authors in ref.~\cite{Chetyrkin:1997vx}
\begin{eqnarray}
\alpha_{em}^{-1}= \alpha^{-1}(\mu_W)  = 130.3\pm 2.3
\label{eq9}
\end{eqnarray}
and used in subsequent studies
\cite{Buras:1997bk,Ciuchini:1997xe}. Since those analyses did not
address QED corrections to $R$, the renormalization of $\alpha(\mu)$
was largely arbitrary. However, a priori, one should expect the fine
structure constant $\alpha=1/137.036$, renormalized at $q^2=0$, to be
more appropriate for real photon emission. Our subsequent loop study
confirms that expectation.

Employing \cite{Chetyrkin:1997vx,Buras:1997bk,Ciuchini:1997xe}
\begin{eqnarray}
{|V_{ts}^*V_{tb}|\over |V_{cb}|} &=& 0.976\pm 0.010,
\nonumber \\
{m_c({\rm pole}) \over m_b({\rm pole})} &=& 0.29 \to
g(m_c^2/m_b^2) = 0.542,
\label{eq10}
\\
F&=& 0.926, \qquad |D^{\rm eff}|=0.373, \nonumber
\end{eqnarray}
the value of $\alpha_{em}$ in (\ref{eq9}) and error analysis in
ref.~\cite{Buras:1997bk,Ciuchini:1997xe} lead to 
\begin{eqnarray}
R^{\rm QCD}&=& 3.32 \pm 0.27\times 10^{-3},
\nonumber\\
{\cal B}(B\to X_s\gamma) &=& 3.58\pm 0.33\times 10^{-4},
\label{eq11}
\end{eqnarray}
which differ from the experimental results by about $1.4\sim 
1.6\sigma$.

Comparison of $R^{\rm exp}$ and $R^{\rm QCD}$ has been used to
constrain or suggest the presence of ``new physics.''  It is a
particularly sensitive probe of supersymmetry
\cite{Hewett:1997ct,Garisto:1993jc} and charged Higgs loops
\cite{Hewett93,Ciuchini:1997xe,Aliev}.  Future high statistics
$\Upsilon(4S)$ studies are expected to further improve $R^{\rm exp}$
and lead to an interesting confrontation with supersymmetry
expectations. Indeed, one can anticipate the error on $R^{\rm exp}$ to
be reduced to about $\pm 10\%$ and in the long term, perhaps $\pm 5\%$
is feasible. It is, therefore, important to fine tune the theoretical
prediction for $R$ as much as possible. With that goal in mind, we
examine here the ${\cal O}(\alpha)$ electroweak corrections to $R$.

\begin{figure} 
\hspace*{-33mm}
\begin{minipage}{16.cm}
\vspace*{-3mm}
\[
\hspace*{-16mm}
\mbox{ 
\begin{tabular}{cc}
\psfig{figure=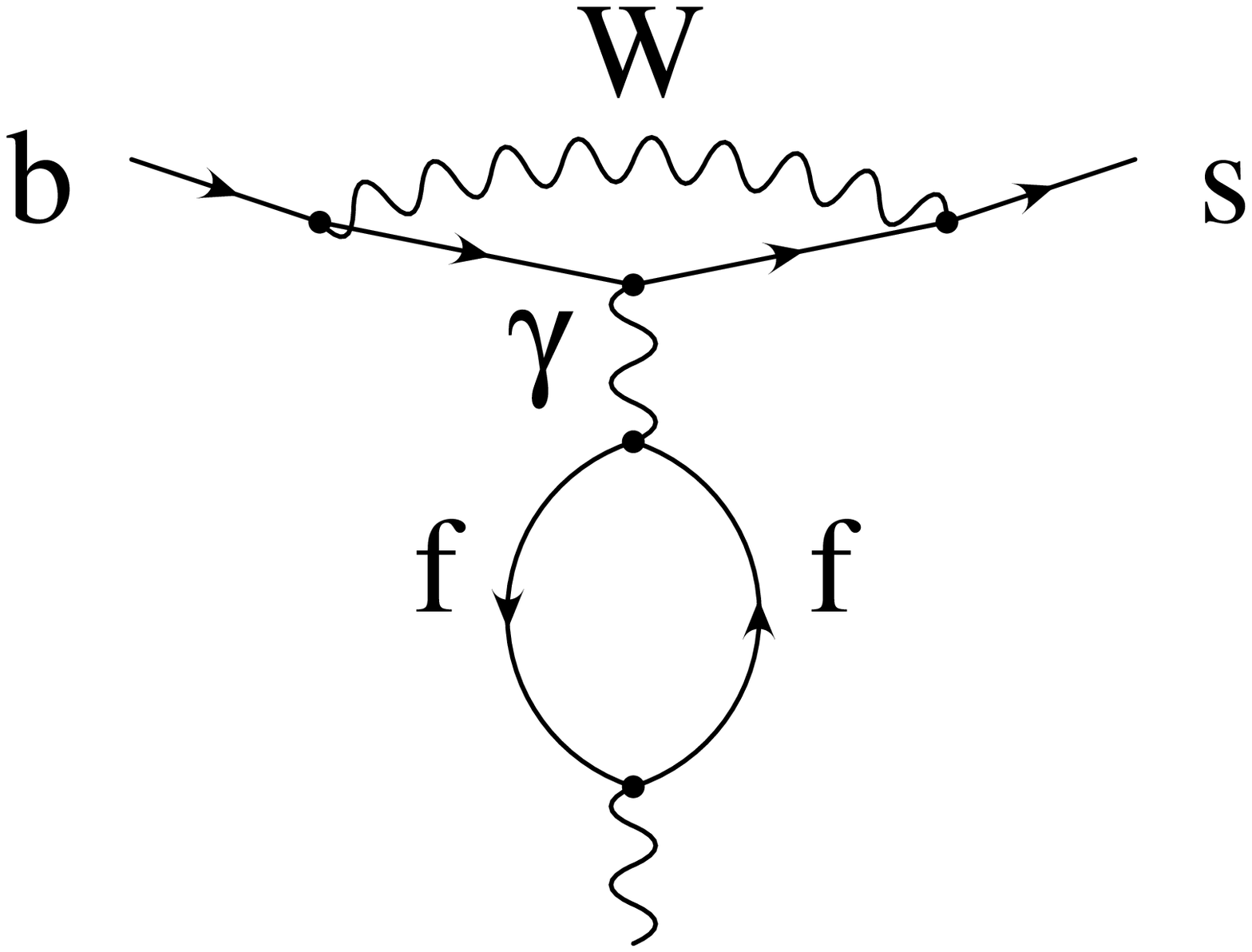,width=38mm,bbllx=72pt,bblly=250pt,%
bburx=544pt,bbury=542pt} 
&\hspace*{13mm}
\psfig{figure=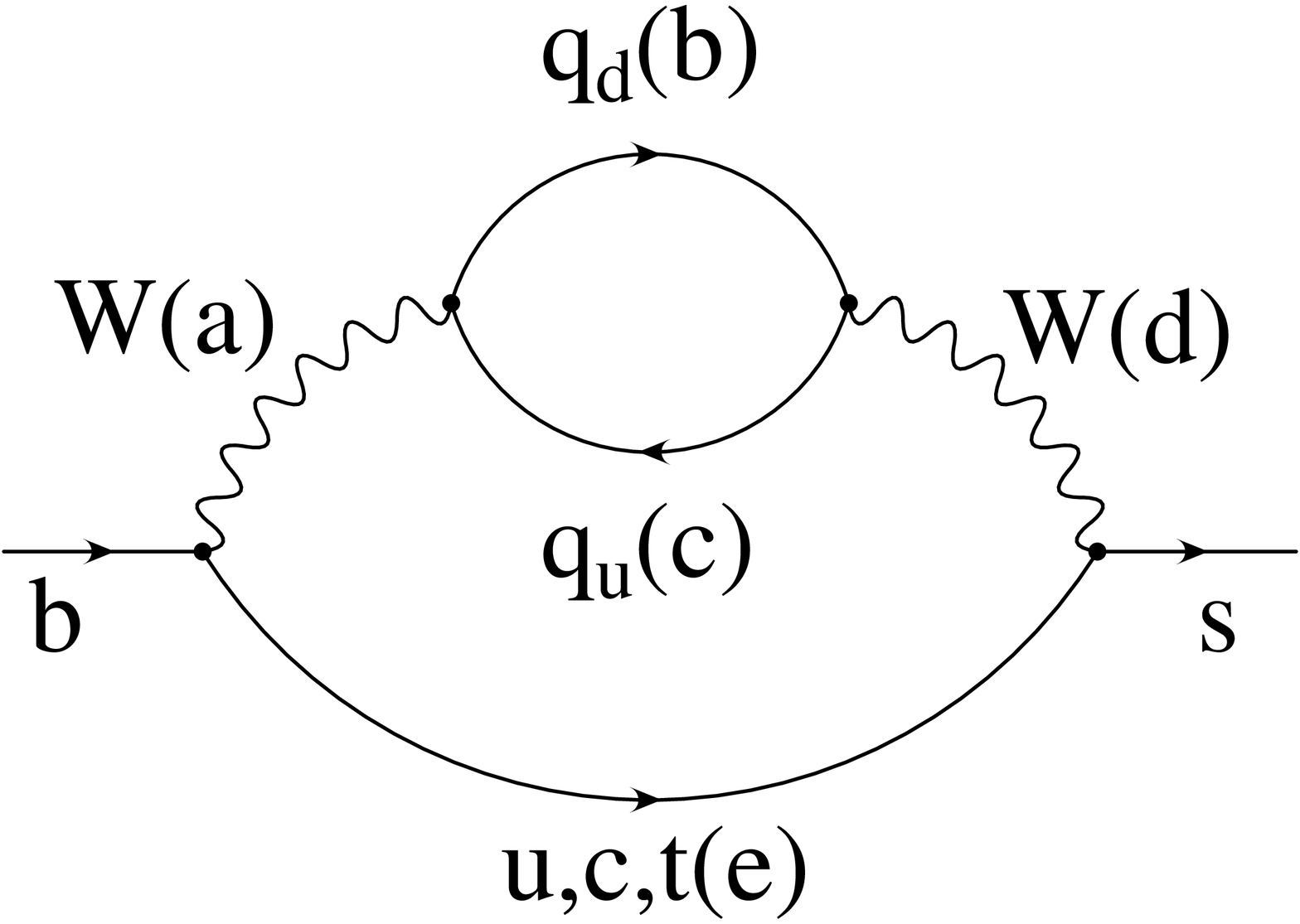,width=30mm,bbllx=210pt,bblly=260pt,%
bburx=630pt,bbury=400pt}    
\\[4mm]
(a) & (b)
\end{tabular}
}
\]
\end{minipage}
\caption{(a) An example of vacuum polarization renormalization of
$\alpha$ by the fermion loops. (b) Fermionic loop corrections to $b\to
s\gamma$.  Letters in parentheses label distinct lines from which the
photon can be emitted.  There is also a contribution from leptons in
the $W$ propagator loop.}
\label{fig2}
\end{figure}

\begin{figure} 
\hspace*{-37mm}
\begin{minipage}{16.cm}
\vspace*{3mm}
\[
\hspace*{-6mm}
\mbox{ 
\begin{tabular}{cc}
\psfig{figure=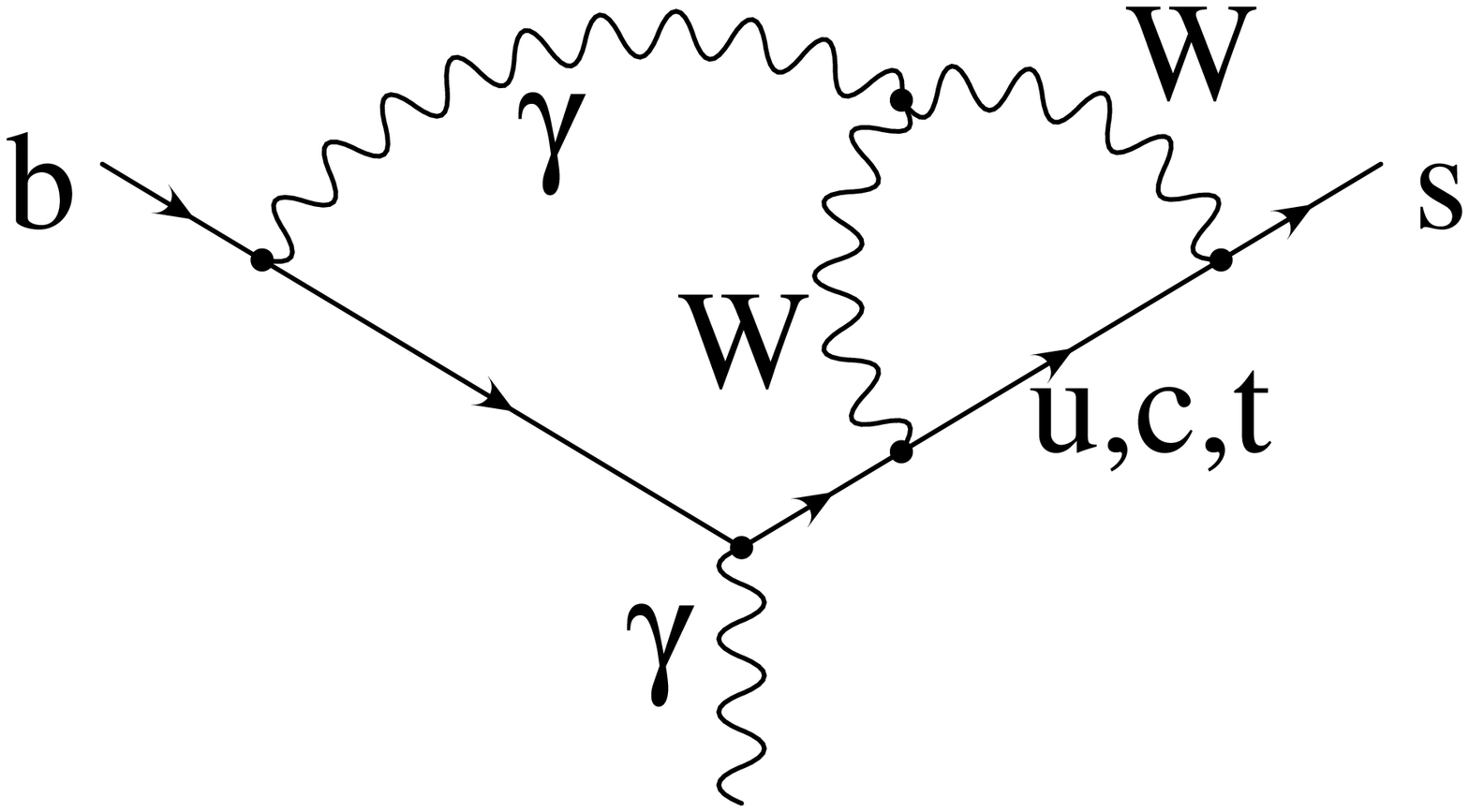,width=35mm,bbllx=72pt,bblly=291pt,%
bburx=544pt,bbury=520pt} 
& \hspace*{5mm}
\psfig{figure=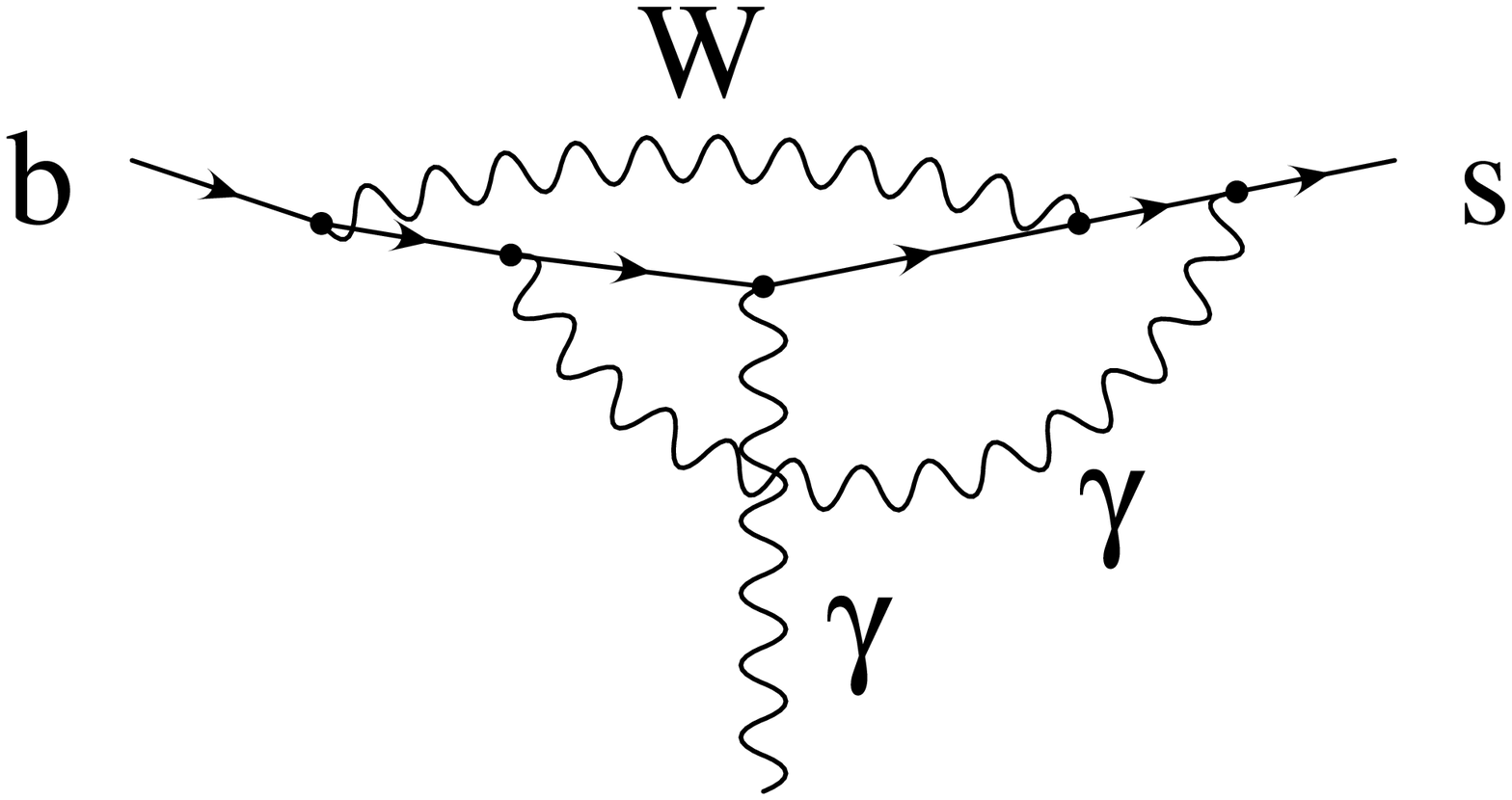,width=35mm,bbllx=72pt,bblly=291pt,%
bburx=544pt,bbury=520pt} 
\end{tabular}
}
\]
\end{minipage}
\caption{Examples of the two-loop diagrams where a virtual photon
exchange gives a short-distance logarithmic contribution.}
\label{fig3}
\end{figure}

\begin{figure} 
\hspace*{-40mm}
\begin{minipage}{16.cm}
\vspace*{3mm}
\[
\hspace*{-5mm}
\mbox{ 
\begin{tabular}{cc}
\psfig{figure=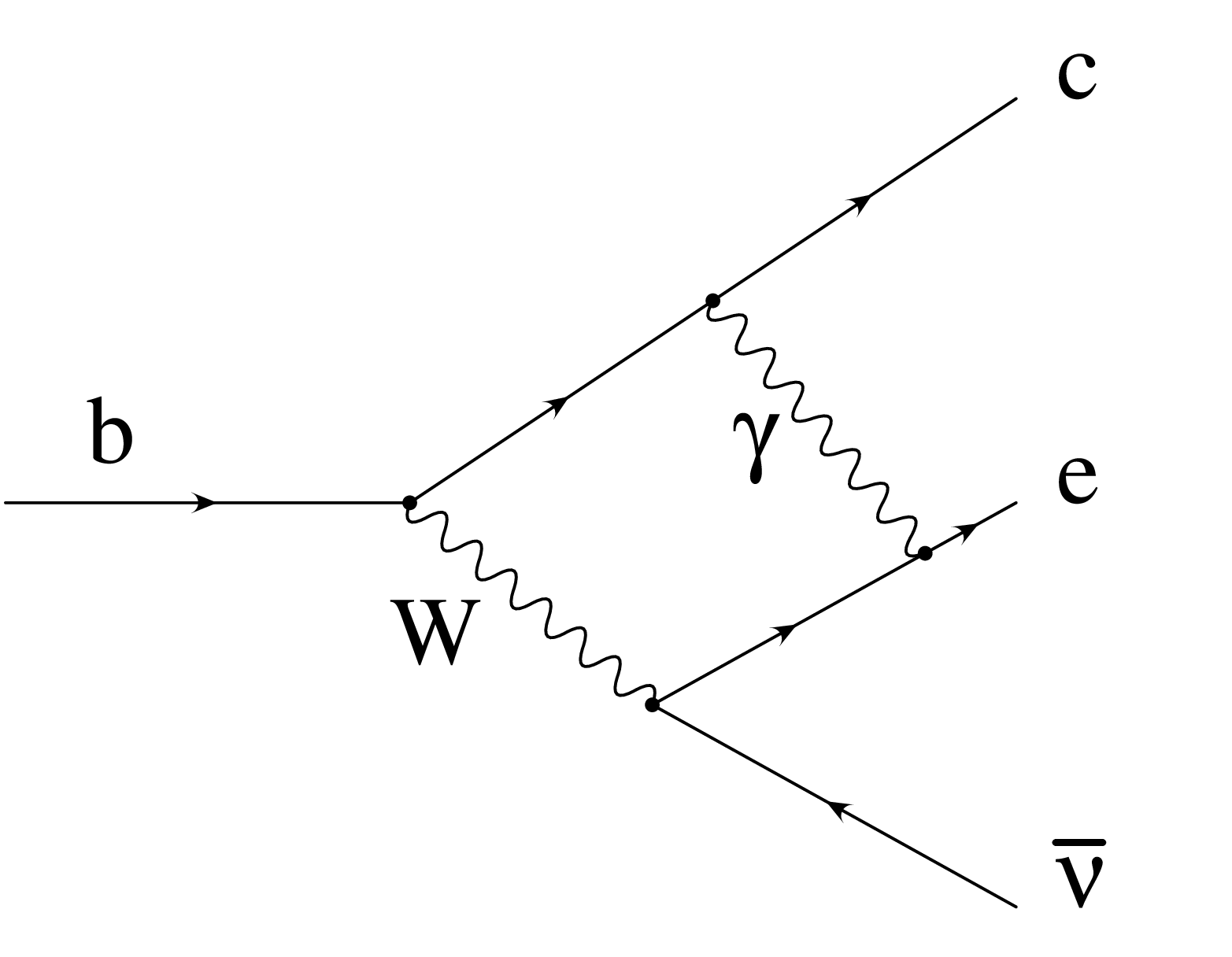,width=30mm,bbllx=72pt,bblly=291pt,%
bburx=544pt,bbury=530pt} 
& \hspace*{10mm}
\psfig{figure=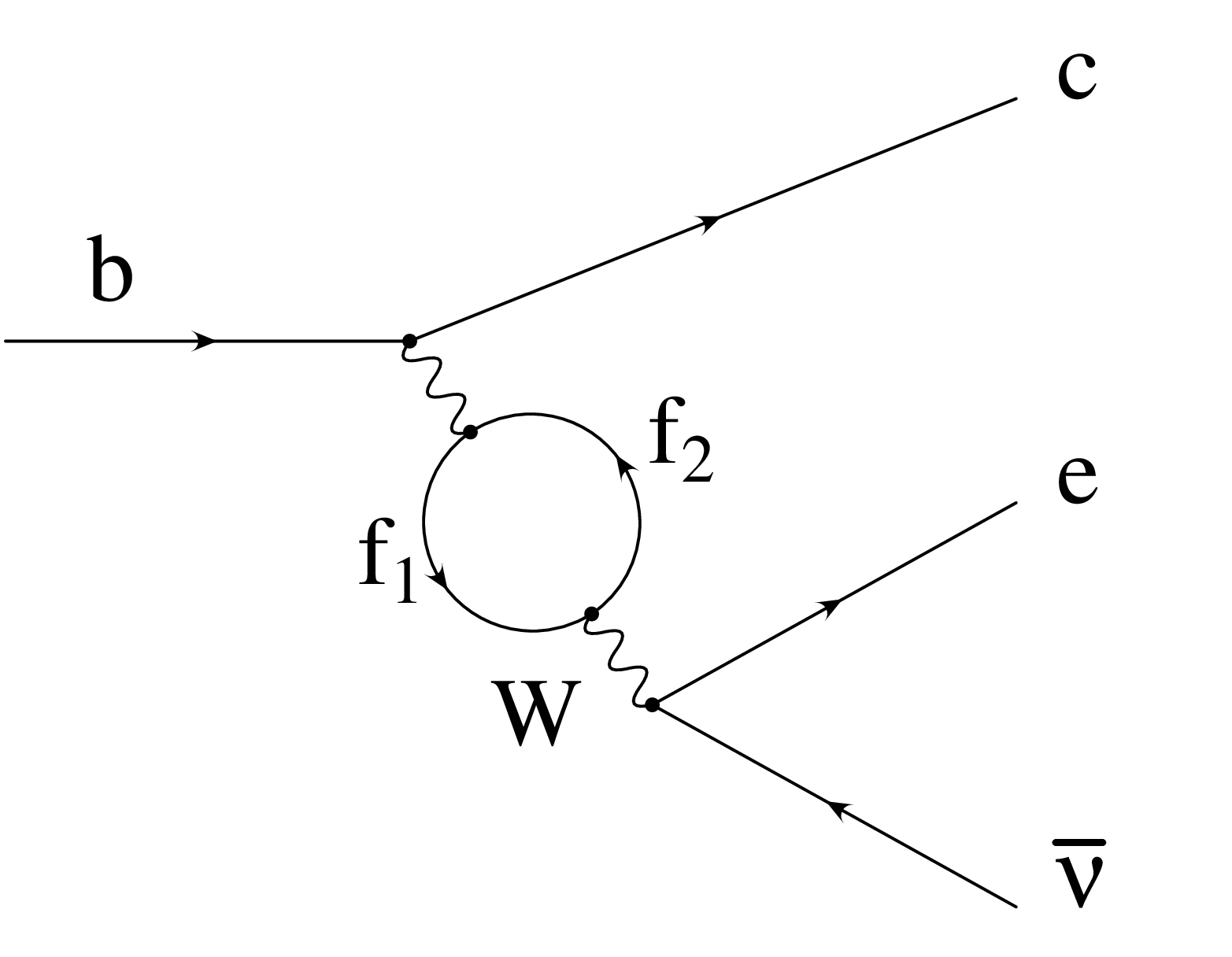,width=30mm,bbllx=72pt,bblly=291pt,%
bburx=544pt,bbury=530pt} 
\end{tabular}
}
\]\vspace*{10mm}
\end{minipage}
\caption{Examples of electroweak corrections to the decay $b\to ce\nu$.}
\label{fig4}
\end{figure}

A study of the ${\cal O}(\alpha)$ corrections to $R$ must entail two
loop contributions to $\Gamma(b\to s\gamma)$ as well as one loop
corrections to $\Gamma(b\to ce\nu)$. A complete calculation of all
contributions would be very difficult and would have to confront long
distance hadronic uncertainties as well as experimental acceptance
conditions. Such a thorough undertaking is not yet warranted by the
data and may not even be required at the $\pm 5\%$ level. It is,
however, important to consider potentially large effects and estimate
their modification of $R$. Here, we examine two such classes of
corrections: contributions from fermion loops in gauge boson
propagators ($\gamma$ and $W$) and short-distance photonic loop
corrections. Examples of those effects for $b\to s\gamma$ are
illustrated in figs.~\ref{fig2} and~\ref{fig3}. Analogous one loop
corrections to $b\to ce\nu$ are pictured in fig.~\ref{fig4}.

We first present our result and then comment on the origin and
magnitude of the various corrections. We find $R^{\rm theory}=R^{\rm
QCD+EW}$ is given by
\begin{eqnarray}
\lefteqn{R^{\rm theory}= R^{\rm QCD}}\nonumber \\
&\times &
\left[{ \alpha \over \alpha_{em}}\right]
\left[ 1-{1\over |D^{\rm eff}|} {\alpha_{em}\over \pi} \left(
{1\over s_W^2} f\left({m_t^2\over m_W^2}\right) 
-{8\over 9}C_7^0\left( {m_t^2\over m_W^2}\right) 
\ln\left({m_W\over m_b}\right) 
+{104\over 243} \ln\left({m_W\over m_b}\right) 
\right)\right]
\nonumber \\
&\times&
\left[ 1-{2\alpha_{em}\over \pi} \ln\left({m_Z\over m_b}\right) \right]
\label{eq12}
\end{eqnarray}
where $\alpha_{em}$ and $|D^{\rm eff}|$ can be found in (\ref{eq9}) and
(\ref{eq10}) while
\begin{eqnarray}
\alpha^{-1}&=&137.036, \qquad s_W^2\equiv \sin^2\theta_W \simeq 0.23,
\nonumber \\
 f(x)&=& \frac{\pi^2(2-3x) (2-x) x}{48} 
+  \frac{x\left( 176 - 373x + 491{x^2} - 468{x^3} + 72{x^4} \right) }
    {192\left( 1 - x \right)^3}
\nonumber \\
&+&   \frac{(2-x) x^2 \left( 5 - 14 x + 11 x^2 - 3 x^3 \right)}
      {8(1-x)^3} {\rm Li_2} \left(1 - \frac{1}{x}\right)
\nonumber \\
&-&\frac{x (2+x)\left( 7 + 16 x - 47 x^2 \right) }{96(1-x)^2}\ln(x-1)
\nonumber \\
&+&\frac{x\left(80-115x+200x^2-425x^3+220x^4-83x^5 \right) }{96(1-x)^4}\ln x
\nonumber \\
&+&{x^2(4-5x-3x^2)\over 16(1-x)^3}\ln x
\left[{5-3x+x^3\over (1-x)^2}\ln x- (2+x)\ln(x-1)\right]
\simeq 0.77,
\label{eq13} \\[1mm]
C_7^0(x) &=& {3x^3-2x^2\over 4(x-1)^4}\ln x 
- {8x^3+5x^2-7x\over 24(x-1)^3}\simeq-0.19 \qquad \mbox{(for
$\overline{m}_t = 167$ GeV).}
\nonumber
\end{eqnarray}
The first correction factor in (\ref{eq12}) 
\begin{eqnarray}
{ \alpha \over \alpha_{em}} = {130.3\over 137.036} = 0.951
\end{eqnarray}
is due to electric charge renormalization, illustrated in
fig.~\ref{fig2}(a). Fermion loops decrease the value of
$\alpha_{em}(\mu)$ by 5\% in going from the short-distance
renormalization condition of ref.~\cite{Chetyrkin:1997vx} to the more
appropriate $q^2=0$ physical condition. This effect is somewhat
trivial in origin, but does represent an important 5\% reduction
neglected in recent analyses. Note, that unlike $\alpha_{em}$, the
physical fine structure constant has essentially no uncertainty.

The $f(m_t^2/m_W^2)$ term in (\ref{eq12}) comes from fermion loop
corrections in the $W$ propagator and $WW\gamma$ vertex as illustrated
in figs.~\ref{fig2}(b) and \ref{fig4}. It leads to a decrease in $R$
by about 2.2\%.  Three loop QCD effects reduce that contribution
by a factor $(\alpha_s(m_W)/ \alpha_s(m_b))^{16/23} \simeq 0.7$
\cite{Shifman:1978de}.

The $\ln(m_W/m_b)$ terms in (\ref{eq12}) originate from photonic
corrections to $b\to s\gamma$ of the type illustrated in
fig.~\ref{fig3}.  In the effective theory language, the coefficients
of those logs are given by the anomalous dimensions $\gamma_{77}$ and
$\gamma_{27}$, of which the latter is more difficult to compute, but
can be obtained from analogous QCD calculations
\cite{Grinstein:1990tj,Ciuchini:1994fk} using the translation
$\alpha_s\to -\alpha/6$. Taken together, those terms reduce $R$ by
about 1\%.  Higher order leading QCD corrections decrease that result
by a factor of $\sim 0.55$ \cite{Kagan}.

The final $\ln(m_Z/m_b)$ correction in (\ref{eq12}) stems from
short-distance photonic corrections to $b\to ce\nu$ of the type
illustrated in fig.~\ref{fig4}. That type of correction is generic to
all semileptonic charged current decays \cite{Sirlin82}. It is
generally factored out of semileptonic $B$ decays before the
extraction of $|V_{cb}|$; hence, it must be explicitly included here
\cite{Atwood90}. That effect reduces $R$ by about 1.4\%.

All of the above corrections have the same sign and when taken
together reduce $R$ by about 8.3\%. There are other EW radiative
corrections of ${\cal O}(\alpha/\pi)$ and ${\cal O}(G_\mu
m_t^2/8\sqrt{2}\pi^2)$ \cite{Strumia} as well as additional 3 loop QCD
effects which we have not computed. We have examined some of those
corrections and found typically $\alt {\cal O}(0.5\%)$ contributions
of both signs.  In addition, long-distance QED effects including two
photon radiation could be several percent.  For example, in
ref.~\cite{Atwood90} it was shown that final state Coulomb
interactions are likely to increase $\Gamma(B^0 \to X_c e\nu)$ by
about 2\% relative to $\Gamma(B^+ \to X_c e\nu)$. However, currently
such corrections may be effectively absorbed in $|V_{cb}|$ and its
uncertainty, since experimental studies do not address their presence.
Hence, we do not include them here.  We see no reason for the sum of
neglected effects to be particularly large but, nevertheless, assign a
conservative $\pm 2\%$ uncertainty to them.  In that way, we find
\begin{eqnarray}
R^{\rm theory}=R^{\rm QCD + EW}\simeq (0.917\pm 0.020) R^{\rm QCD}.
\label{eq15}
\end{eqnarray}
Since $R^{\rm QCD}$ previously contained a $\pm 1.8\%$ uncertainty due
to $\alpha_{em}$ \cite{Chetyrkin:1997vx} which our analysis does not
have, the overall theoretical uncertainty in $R$ remains essentially
unchanged.  From (\ref{eq15}) and (\ref{eq11}), we find the reduced
predictions
\begin{eqnarray}
R^{\rm theory} &=& 3.04\pm 0.25\times 10^{-3},
\nonumber \\
{\cal B}(B\to X_s\gamma)^{\rm theory} &=& 3.28 \pm 0.30\times 10^{-4},
\label{eq16}
\end{eqnarray}
which are in better agreement with experiment.

Rather than using the value of $|V_{ts}^*V_{tb}| / |V_{cb}|$ in
(\ref{eq10}), we can employ 3 generation unitarity which implies
\begin{eqnarray}
{|V_{ts}^*V_{tb}|^2 \over |V_{cb}|^2} = |V_{tb}|^2\left(
1- {|V_{td}|^2 - |V_{ub}|^2 \over |V_{cb}|^2} \right)
\label{eq17}
\end{eqnarray}
or using $|V_{cb}|=0.039\pm0.003$, and the Wolfenstein parameterization
\cite{Wolfenstein:1983yz,Rosner:1997gb} of the CKM matrix
\begin{eqnarray}
{|V_{ts}^*V_{tb}|^2 \over |V_{cb}|^2} = 0.950(1+0.10\rho).
\label{eq18}
\end{eqnarray}
Constraints from $b\to ue\nu$, $B^0-\bar B^0$ oscillations and
$K^0-\bar K^0$ mixing currently limit $\rho$ to \cite{Rosner:1997gb} 
\begin{eqnarray}
\rho \simeq 0.03\pm 0.24,
\label{eq19}
\end{eqnarray}
and the corresponding range for $|V_{ts}^*V_{tb}|/ |V_{cb}|$ in
(\ref{eq10}). However, keeping $\rho$ arbitrary, we find
\begin{eqnarray}
R^{\rm theory} = 3.03 \pm 0.25 \times 10^{-3} (1+0.10\rho),
\label{eq20}
\end{eqnarray}
which on comparison with (\ref{eq7}) implies
\begin{eqnarray}
\rho=-1.7\pm 1.9.
\end{eqnarray}
That result is consistent with (\ref{eq19}).  Future $\pm 5\%$
measurements of $R^{\rm exp}$ would determine $\rho$ to $\pm
0.5$. Although not competitive with other methods for pinpointing
$\rho$, such studies would be very valuable for constraining or
providing evidence for ``new physics.''

In summary, we have found a $8\pm 2 \% $ reduction in the Standard
Model prediction for $R^{\rm theory}$ due to EW radiative
corrections. That result improves agreement with experiment,
\begin{eqnarray}
R^{\rm exp}/R^{\rm theory} = 0.83\pm 0.18.
\end{eqnarray}
It can be used to constrain ``new physics'' effects such as
supersymmetry; however, those studies are beyond the scope of this
paper. Further reduction in the $\pm 8\%$ uncertainty of $R^{\rm
theory}$ will require a better determination of $m_c({\rm pole}) /
m_b({\rm pole})$ (currently the main theoretical uncertainty), perhaps
by studies of semileptonic $B$ decay spectra.  This uncertainty could
also be reduced by re-analyzing QCD corrections to $b\to s\gamma$
using low-scale running quark masses \cite{Uraltsev95upset}.

We look forward to future improved measurements of ${\cal B} (B\to
X_s\gamma)$ and ${\cal B} (B\to X_c e\nu)$ and their confrontation
with theory.

\vspace*{.3cm} A.C. thanks K. Melnikov for collaboration at an early
stage of this project and M. Misiak for helpful remarks.  This work
was supported by the DOE contract DE-AC02-98CH10886.


\end{document}